\pdfoutput=1
\documentclass[journal=jceda8,manuscript=article]{achemso}

\usepackage{chemformula} 
\usepackage[T1]{fontenc} 
\usepackage{dialogue,siunitx}
\usepackage{etoolbox}
\usepackage{bm}
\DeclareSIUnit\molar{\textsc{m}}
\DeclareSIUnit{\wtpercent}{wt\%}
\makeatletter
\appto{\PreDialogue}{\global\@newlistfalse}
\makeatother

\author{Laszlo Frazer}
\affiliation[Unknown University] {ARC Centre of Excellence in Exciton Science and School of Chemistry, Monash University, Clayton, Australia}
\author{Heather F. Higginbotham}
\affiliation[Unknown University] {School of Chemistry, Monash University, Clayton, Australia}
\author{Toby D. M. Bell}
\affiliation[Unknown University] {School of Chemistry, Monash University, Clayton, Australia}
\email{toby.bell@monash.edu}
\author{Alison M. Funston}
\affiliation[Unknown University] {ARC Centre of Excellence in Exciton Science and School of Chemistry, Monash University, Clayton, Australia}
\email{alison.funston@monash.edu}

\title[An \textsf{achemso} demo]{`It's fundamental': Quantum dot blinking experiment to teach critical thinking}

\keywords{Upper-Division Undergraduate, Laboratory Instruction, Problem Solving / Decision Making, Semiconductors, Fluorescence Spectroscopy, Nanotechnology}

\begin{document}

\begin{tocentry}
\centering
  \includegraphics[height=35mm]{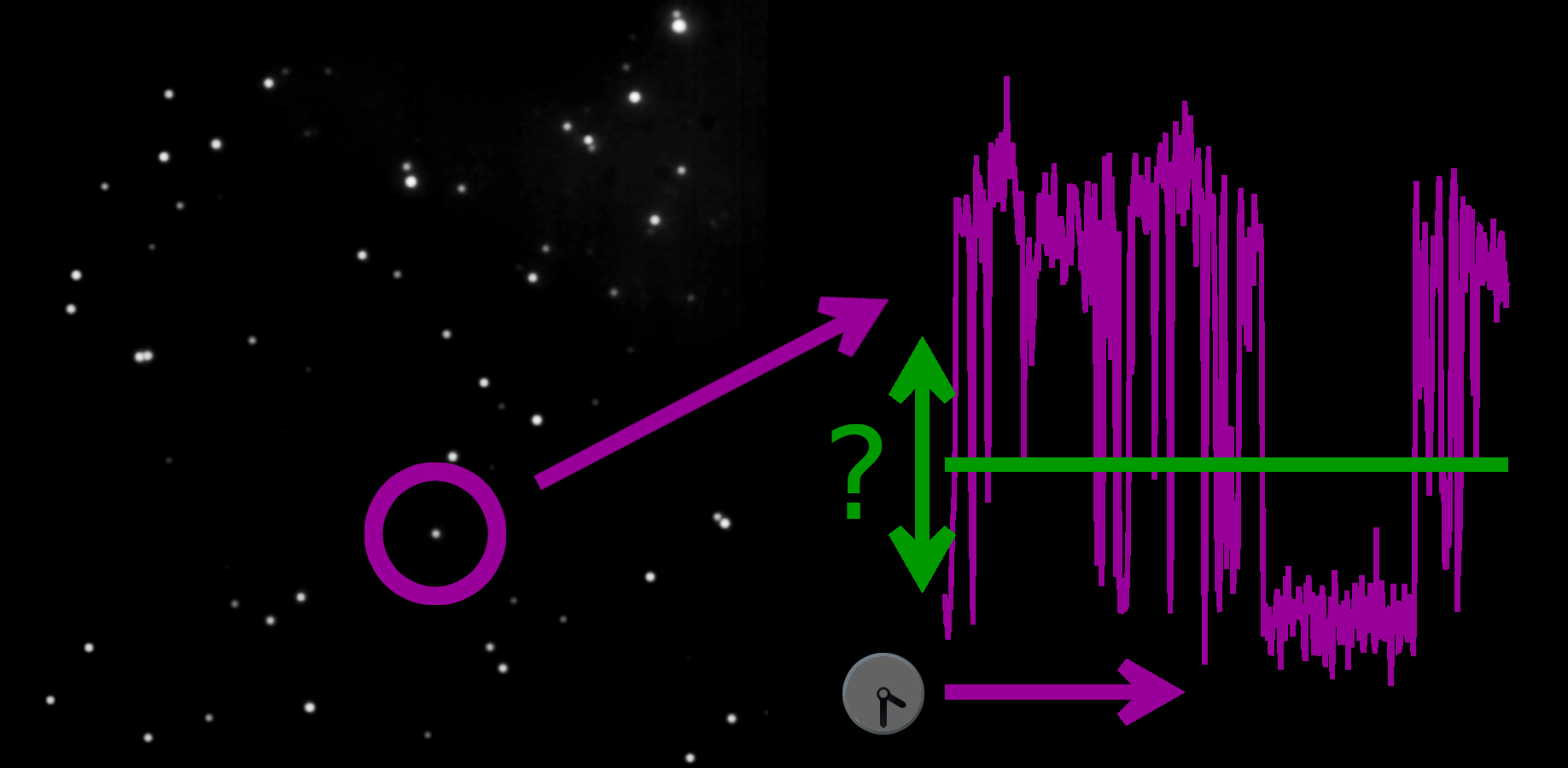}


\end{tocentry}

\begin{abstract}
  Analysis of stochastic processes can be used to engender critical thinking.  Quantum dots have a reversible, stochastic transition between luminescent and non-luminescent states.  The luminescence intermittency is known as blinking, and is not evident from ensemble measurements.  In order to stimulate critical thinking, students design, perform, and analyze a semiconductor quantum dot blinking laboratory experiment.  The design of the experiment and stochastic nature of the data collected require students to make judgements throughout the course of the single-particle measurement and analysis. Some of the decisions do not have uniquely correct answers, challenging the students to engage in critical thinking.  We propose that students' self-examined decision making develops a constructivist view of science.  The experiment is visually striking, interdisciplinary, and develops higher order thinking.
\end{abstract}

\section{Introduction}
Critical thinking is an important skill for students to acquire.\cite{kogut1996critical,warren2018quantitative} While the acquisition of critical thinking skills is often given as a goal of laboratory instruction, evidence indicates it is seldom achieved\cite{holmes2015teaching} and it has been posited that expository laboratory experiments may not develop critical thinking if they have a predetermined outcome.\cite{domin1999review,schoffstall2007incorporating}  We use the quantum randomness of single-particle measurements to cause students to make judgments about measurement quality.  We describe an advanced experiment that promotes awareness of experimental bias in physical science research.
The educational activity is designed for students at or above the advanced undergraduate level.

There is disagreement about the nature of critical thinking.\cite{danczak2017does,aloisi2018threats} Different types of skills may be included under the `critical thinking' umbrella.  Our investigation focuses on the evaluation aspect of critical thinking.  We set an evaluation\cite{bloom} learning objective for students to judge the quality of their measurements.  As a second objective, we prompt students to analyze\cite{bloom} by identifying the bias that arises from their judgments.  

Evaluation skills may be related to beliefs about the nature of knowledge.  Students are often unaware of subjective influences on science.\cite{ryan1992students}  In this experiment, students perform an experiment that challenges positivist epistemology or epistimic authoritarianism.\cite{raviv2003teachers,hornikx2011epistemic,njegovan2011characteristics,blue2018assessing,conley2004changes}  
When students identify that their judgments cause bias, conceptual conflict occurs between positivist beliefs and experiences supporting constructivist epistemology\cite{bodner1986constructivism,bodner2001many}.
The conflict has the potential to change students' views about the nature of scientific reasoning.


The educational value of single-particle measurements\cite{harbron2002poisson,zimmermann2004fluorescence,ito2008observation} is that they challenge students' belief that the properties of ensembles are identical to the properties of individual particles.\cite{tversky1971belief,rabin2002inference}  This type of belief is reinforced by the wide-spread use of ensemble measurements in chemical education, such as in optical spectroscopy, rheology, or NMR.  However, single-particle measurements can detect rare events that are masked in ensembles.  For example, the discovery of stable isotopes proved that neon atoms are not all identical.\cite{griffiths1997jj}

The apparatus students use in this experiment is primarily used as a research tool for wide field microscopy.\cite{zimmermann2004fluorescence} Lower cost adaptations are possible. The experiment and the surrounding classroom activities introduce students to the diffraction limit\cite{ito2008observation} and the valuable applications of high resolution imaging.\cite{coltharp2012superresolution,huang2009super,schermelleh2010guide,whelan2014focus,whelan2015super}  

Quantum dots are semiconductor nanoparticles.\cite{klimov2010nanocrystal}  They exhibit broad absorption spectra at energies above their bandgap along with narrow photoluminescence.  Quantum dots generally outperform molecular dyes in photostability, and can be investigated in air at room temperature.\cite{zimmermann2004fluorescence}  These properties, along with their inherent polydispersity, make them ideal for accessible single particle measurements at the undergraduate level.  Here, we focus on quantum dot properties resulting from bottom-up colloidal synthesis.  A distinctive property of such nanoparticles is that their attributes, particularly their quantum yield, are highly sensitive to modification of the surface\cite{bullen2006effects} due to their high surface to volume ratio. 

Single quantum dots are well known to blink.\cite{nirmal1996fluorescence,lee2009brightening} 
Blinking occurs when a photoluminescent particle (or molecule) temporarily ceases to emit light.\cite{zimmermann2004fluorescence}  Illuminated quantum dots generally exhibit two-state behaviour, consisting of a brightly luminescent on-state and a dark off-state.  The ground state is ignored by the two-state theoretical framework for blinking.  However, there can also be additional gray states caused by trap states and Auger processes.\cite{yuan2018two}  These have an intermediate brightness, between the brightness of the on-state and off-state. 

The photoluminescence of single quantum dots\cite{riel2008introduction} has properties that can contribute to students' awareness of measurement bias.  The lifetime of the blinking off-state is distributed according to a power law.\cite{shimizu2001blinking,cichos2007power,verberk2002simple,kuno2000nonexponential,kuno2001off,gomez2006blinking,efros2016origin,pelton2004characterizing}  The origin of the power law is a current topic of investigation.  There is no typical duration for an off-state because the distribution of off-state durations has no statistical moments.  If a quantum dot is not detected by measuring photoluminescence, it is impossible to determine if it is absent, incapable of luminescence, or simply in the off-state for a long time.  

The power law probability distribution function cannot be normalized over its entire domain.  There are short off-periods that cannot be resolved and long off-periods whose ends are missed.  The presence of gray states can further complicate the experimenters' decisions about the presence of a particle and criteria for detecting an off-state.  Even in the absence of a gray state, the quantum dot may switch between on- and off-states in the middle of an acquisition, producing a data point that is between on and off.\cite{rabouw2019microsecond}  
When multiple particles are separately measured, the peak brightness of one may be less than the background brightness of another, resulting in a need to discard data or use multiple criteria to detect blinking.  Additionally, photoluminescence observed from a diffraction limited spot, nominally attributed to a single quantum dot, may in fact be due to two (or more) nanocrystals separated by a distance less than the diffraction limit.  

We use a pedagogical design where students make decisions about single quantum dot luminescence measurements.  Then the students analyze the relationship between decisions and results.  The process is an example of constructivism in action.  The action is constructivist because what is learned depends on the students' choice, and not solely on the instructor or on the physics.  The experiment is designed for final year undergraduates, masters students, or first year doctoral students in chemistry, (bio)physics, nanotechnology, materials science, or quantum information.  

\section{Experimental Methods}
A protocol written for a student audience and the survey protocol are included in the Supporting Information.

\subsection{Materials}
\begin{sloppypar}
Toluene (HPLC+), poly(methyl methacrylate) (GPC Standard, $M_W=$ 350,000), hexadecylamine (98\%), and 1-octadecanethiol (98\%) from Sigma-Aldrich were used as received. Ultrapure water (18.2 M$\Omega$, Milli-Q) was used for all the procedures.

CdSe quantum dots \cite{van2005nucleation} and CdSe/CdS/ZnS core-shell-shell quantum dots\cite{van2007review,talapin2004cdse} were synthesized in advance according to literature methods.  

Glass coverslips were cleaned in advance.  The coverslips were soaked in chloroform for 30 minutes, rinsed, then sonicated sequentially in acetone, \SI{1}{\molar} aqueous NaOH and deionized water for 20 minutes each respectively, with extensive rinsing between solvents.\cite{harbron2002poisson} The coverslips were stored in a clean beaker in deionized water until needed. 

A number of pre-cleaned coverslips were rinsed with ultrapure water and dried with a stream of N$_{2}$ in preparation for sample deposition by each student. All glassware including sample vials, dried coverslips and pasteur pipettes were placed in an UV-ozone cleaner for 15 minutes immediately prior to use.

Quantum dots of either type were serially diluted in toluene or \SI{1}{\wtpercent} poly(methyl methacrylate)/toluene solution.
Optionally, \SI{1}{\milli\molar} hexadecylamine or 1-octadecanethiol were included in the solvent during dilution to change the surface chemistry of the quantum dots.  The solution was spin-coated\cite{dabirian2014construction,sadegh2019design} onto pre-cleaned coverslips at 5000 rpm for 60 seconds.  Spin coating is used in single particle imaging because it distributes the particles with a low density.

In the interests of time, each student was limited to preparing and analyzing one solution of quantum dots.  Dilution in pure toluene relies on the presence of the surfactants/ligands from the synthesis to maintain the colloidal stability of the nanoparticles. This particular dilution and its subsequent deposition (via spin coating) on the coverslip should be carried out as quickly as possible to minimise dissociation of the ligands upon dilution as this can reduce the quantum dot luminescence.

\end{sloppypar}

\subsection{Data collection}
\begin{figure}
  \centering
  \includegraphics[width=.6\textwidth]{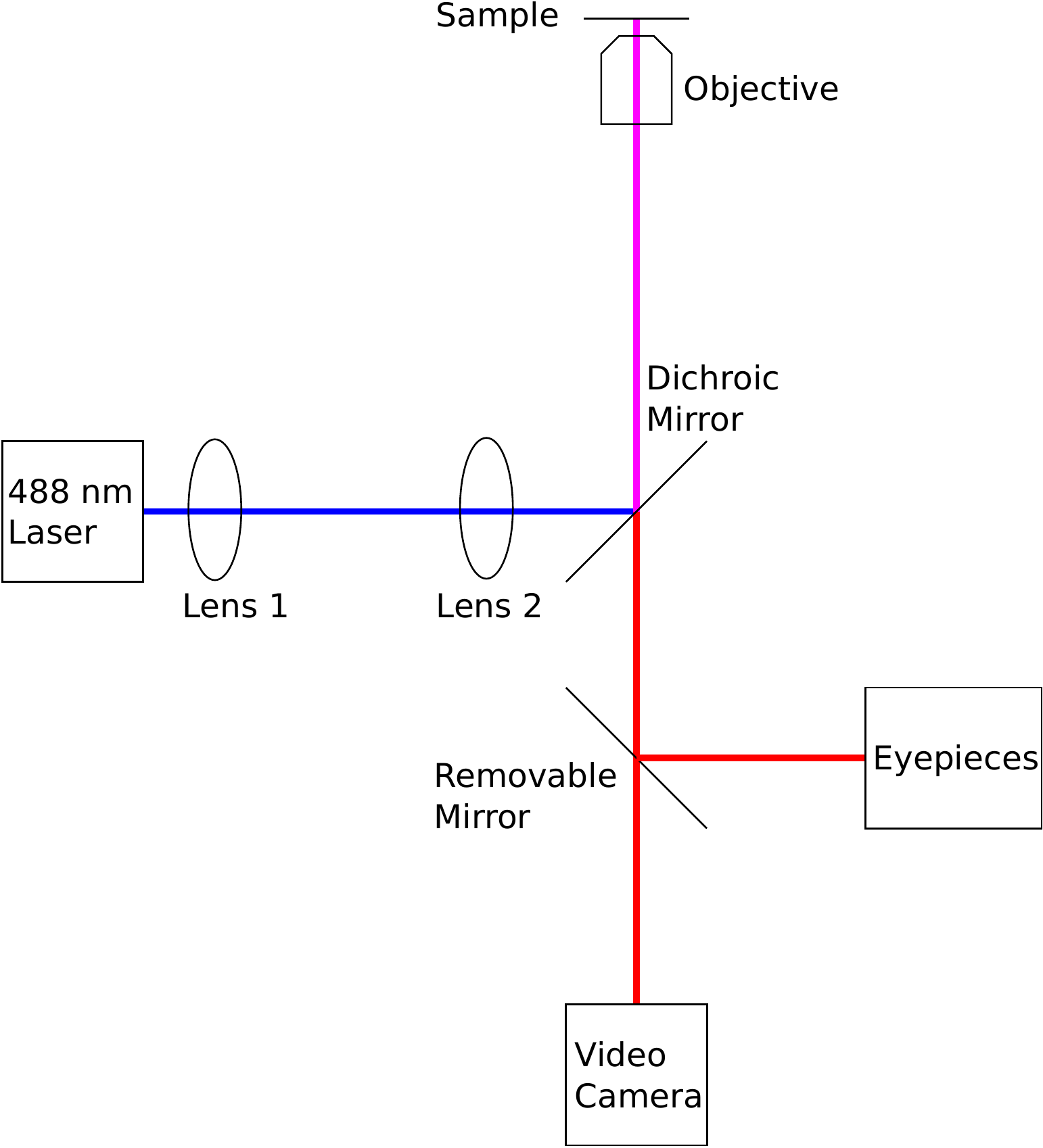}
  \caption{Simplified diagram of the wide-field fluorescence microscope.}
  \label{fig:instrument}
\end{figure}

An inverted microscope (Olympus IX71) set up in a wide-field configuration as illustrated in Fig. \ref{fig:instrument} was used for single particle measurement.  Samples were illuminated using a 488 nm, 200 mW Toptica iBeam Smart 488-S-HP-10901 G0 solid state laser diode.  The laser beam was expanded to a flat field using a lens system and focussed onto the back of the objective.  A 1.49 or 1.4 numerical aperture, $100\times$ magnification objective was used.  A dichroic filter separated the incident laser excitation and the outgoing luminescence.  The luminescence of multiple quantum dots was recorded as a video using an Andor iXon Ultra EMCCD\cite{whelan2014focus} in a darkened room. The luminescence and blinking were readily visible to the dark-adjusted eye through microscope eyepieces. Luminescence blinking was recorded in 100 second, 10 frame per second videos, allowing about five students to complete the experiment per hour.  

\subsection{Hazards}
Lasers may cause eye injury and should not be viewed directly.  Quantum dots may be toxic.\cite{hardman2005toxicologic}  Spin coaters should be guarded and interlocked.  Working in the dark is a trip hazard.

\subsection{Data Analysis}
Using a convenient custom-made data analysis package based on a menu-and-dialogue computer interface, pixels capturing the luminescence from a presumed single quantum dot are defined.  Background areas are also defined.  The analysis software automatically generates a photon trajectory (brightness as a function of time bin, as measured with video frames) from the summed light intensity of all pixels defined as quantum dot luminescence, with background counts from an area of identical size subtracted.  Multiple trajectories are concatenated to increase statistical power easily at the cost of accuracy.  A brightness threshold identifies off-times in the trajectory, from which the software automatically generates a histogram of off-times.  The software reduces a selected domain of durations in the histogram to a power law exponent by log transformation\cite{goldstein2004problems} and Poisson weighted linear regression.  This exponent describes the temporal distribution of blinks.  We do not introduce autocorrelation/power spectral density analysis\cite{verberk2003photon,pelton2004characterizing} or Bayesian estimation,\cite{geordy2019bayesian} which are more complex but do not require a threshold, or change point analysis.\cite{watkins2005detection}  

\subsection{Possible Modifications}
Commercial quantum dots in organic solvents, with luminescence in the visible region and exchangeable surfactants/ligands can be substituted.
Core-shell quantum dots are significantly easier to measure due to their generally higher quantum yields.
Wide field fluorescence instruments\cite{peidle2009inexpensive} are widely available because they are used for epifluorescence imaging in biology.   Quantum dots absorb light over a wide range of the spectrum. 
Illumination may be carried out using visible irradiation at any energy higher than the quantum dot band-gap.   Many lasers or filtered arc lamps provide suitable illumination. 
A large numerical aperture objective is essential to ensure efficient light collection.  Any high quantum efficiency CCD video camera capable of recording at least ten frames per second could be used as a detector.

The surface chemistry of the nanoparticles is manipulated to be either the surfactants/ligands present in a standard colloidal synthesis of CdSe core-shell nanoparticles, alkylamine (primary amine) functionalised or alkanethiol functionalised. The latter two conditions are achieved by a straightforward dilution of the nanoparticles in solutions containing an excess (1 mM) of the ligand. The alkyl chain lengths of the ligands are relatively unimportant provided colloidal stability of the nanoparticles is maintained (generally true for C8 or longer alkyl chains), and so substitutions of these ligands with those containing different alkyl chain lengths is possible. Substitution for secondary amines is also possible. Care should be taken to ensure the ligand does not introduce impurities which fluoresce.

\section{Pedagogical Design}
\begin{figure}
  \centering
  \includegraphics[width=\textwidth]{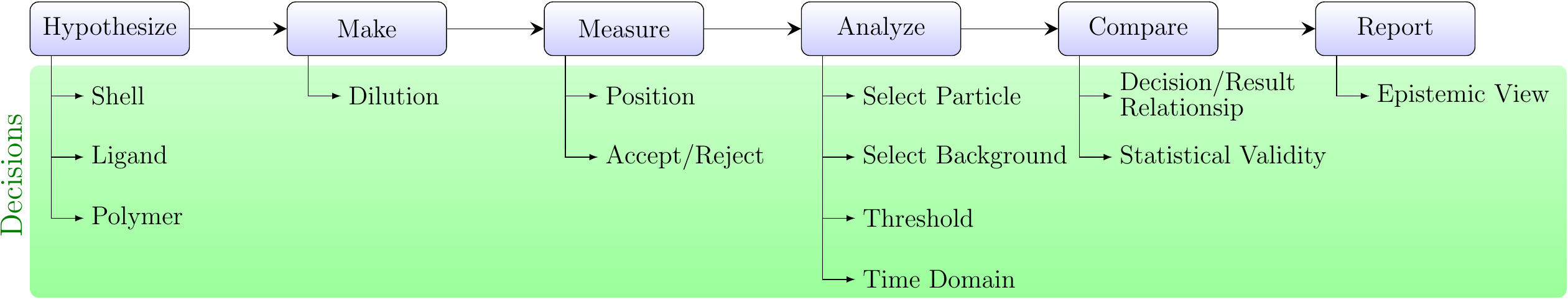}
  \caption{Students perform a series of steps.  For each step, we give examples of decisions students could make. Decisions are opportunities for spontaneous or prompted evaluation and analysis. In other words, they are opportunities for critical thinking.}
  \label{fig:design}
\end{figure}

The experiment is organized as a conventional inquiry-based instructional activity, as illustrated by Fig. \ref{fig:design}.  However, student decision making is not limited to the initial hypothesis-formation step.  Instead, as shown in the green box, opportunities for decision making are interspersed throughout the task.  We give some examples of the most interesting decisions below.

Prior to undertaking the laboratory practical, students have attended lectures introducing them to the phenomenon of quantum dots. The topics covered include quantum confinement, colloidal stability and the passivation of quantum dots via ligands/surfactants or shelling with a wider bandgap material on the (ensemble) quantum yield for quantum dots. Reference papers for these effects are provided throughout the lectures. Students are also directed to consult the Supporting Information and its references to inform themselves before the activity.  

To initiate inquiry based learning,\cite{healey2005linking,savery2015overview,cummins2004prompted,deters2005student,sanger2007effects,cords2012teaching} students form a hypothesis relating the choice of quantum dot type, inclusion of polymer, and choice or omission of ligand on the power law exponent.  These low cost options allow students to explore the factors that determine the blinking and luminescence quantum yield of quantum dots.\cite{gomez2006blinking} 
During sample preparation, students' choice of a sufficiently low quantum dot concentration is essential to achieving the measurement of single quantum dots. Whilst very
general guidance regarding the concentrations required is given, the concentration tolerance of the stock solution to form a sample on which single particle measurements can be carried out is high. The students are required to
decide the actual concentration of the stock solution to spin coat following a prelaboratory discussion considering the diffraction limit,  and processes involved in spin coating. It is not unusual for the students to make a few attempts to prepare a sample with appropriate dispersion, allowing them to appreciate the diffraction limit and particle density requirements for single particle imaging.
The measurement portion of the activity is primarily expository because of time and safety constraints.
While measuring fluorescence, students select a region of the coverslip to measure and may decide to reject their sample preparation in favor of a new sample preparation design.

Students make a series of judgments to reduce a video file to a measurement of the power law exponent for the quantum dot off-state duration.  First, they select pixels in the video that contain a single quantum dot.  Here, students are judging the number of particles present in these pixels, as well as which `single' quantum dots should be selected for measurement in their analysis.  Next, students select the pixels that they use to measure the background brightness.  The background brightness can be time- and space-dependent, leading to the possibility that students may inadvertently select pixels corresponding to a dim quantum dot or a background of different average brightness compared to the quantum dot measurement. They need to decide if the background area is appropriate.  

The students decide a brightness threshold to set, which is then used to calculate the off-times (duration continuously below threshold) of the quantum dots. This decision is complicated by the concatenation of data from different single quantum dots which may have different blinking and/or background brightnesses.  Usually, this decision involves ambiguity because of the stochastic nature of blinking.  Students select a domain of durations within the resulting histogram to exclude invalid data before calculating the power law exponent.  The optimal domain is ambiguous because the onset of duration-dependent errors, such as blinks that continue beyond the end of the video, is gradual.  Care is taken to explain the properties of histograms to the students owing to the difficulty students face understanding histograms.\cite{kaplan2014investigating,cooper2008students,meletiou2005exploring,meletiou2002student,lee2003some} 
Measurement bias inherent to quantum dot properties is relevant to students' decisions.  Students may fail to measure a quantum dot which happens to be off for longer than the experiment.  When a student inadvertently selects two quantum dots, an off state will only be measured when both quantum dots are in the off state.  Then off state durations are underestimated. When selecting a threshold, students may misclassify gray states.  Since the power law probability distribution function is not normalizable, the measurements are inevitably biased by the students' chosen time domain.  These types of bias are example opportunities for students to achieve the evaluation learning goal.

Students test their hypotheses using their own together with their classmates' results.  They may decide upon decision-result relationships and statistical validity at this point.
In their laboratory reports, students are instructed to compare their measurements with their hypothesis about the samples.  Subsequently, they describe how they made judgments about the data.  

Finally, students were prompted to consider why measurements might vary.  Students can choose to support different epistemic views.  These are illustrated in Table \ref{tab:views}.  The pedagogical design emphasizes the constructivist aspects of science.
\begin{table}
    \centering
\begin{tabular}{ll}
Epistemic View&Illustrative Example\\\hline
Authoritarian       & The expert's measurement is more accurate. \\
Positivist          & The measurements are inherently random. \\
Constructivist      & The experimenter's choice of analysis method changed the results.
\end{tabular}
    \caption{Perspectives on the epistemology of the experiment.  Each example illustrates a students' belief about the causes of variance.}
    \label{tab:views}
\end{table}

In our implementation, the experiment is targeted at students who are transitioning to being researchers.  The experience gained serves as an introduction to optical research.  It also informs students about aspects of the interdisciplinary research fields of single-particle spectroscopy, super-resolution microscopy, nanotechnology, quantum information,\cite{pearson2010hands,dehlinger2002entangled,galvez2005interference} and excitonics.  

We implemented the experiment in an honours course at a research intensive Australian university.  An honours course is a fourth year of tertiary education completed after a three year bachelor's degree.  It can be used as a prerequisite for enrollment in doctoral study.   
The students enrolled have an average grade of 70\% or better from undergraduate chemistry studies.  
Honours students complete a nine month capstone project and coursework.  In our context, enabling progression to doctoral research is an objective of honours education. About ten students perform the experiment per year. 

\section{Quantum Dot Blinking Results}

\begin{figure}
  \centering
  \includegraphics[width=.3\textwidth]{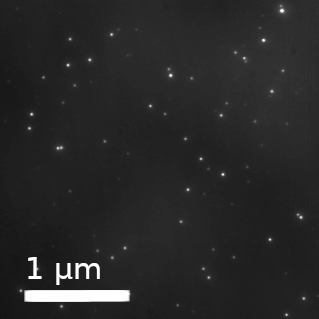}
  \caption{An example of a student-recorded image of quantum dot luminescence.  The experiment is visually striking because students can view the blinking of single quantum dots by eye.}
  \label{fig:blink}
\end{figure}

\begin{figure}
    \centering
    \includegraphics[width=\textwidth]{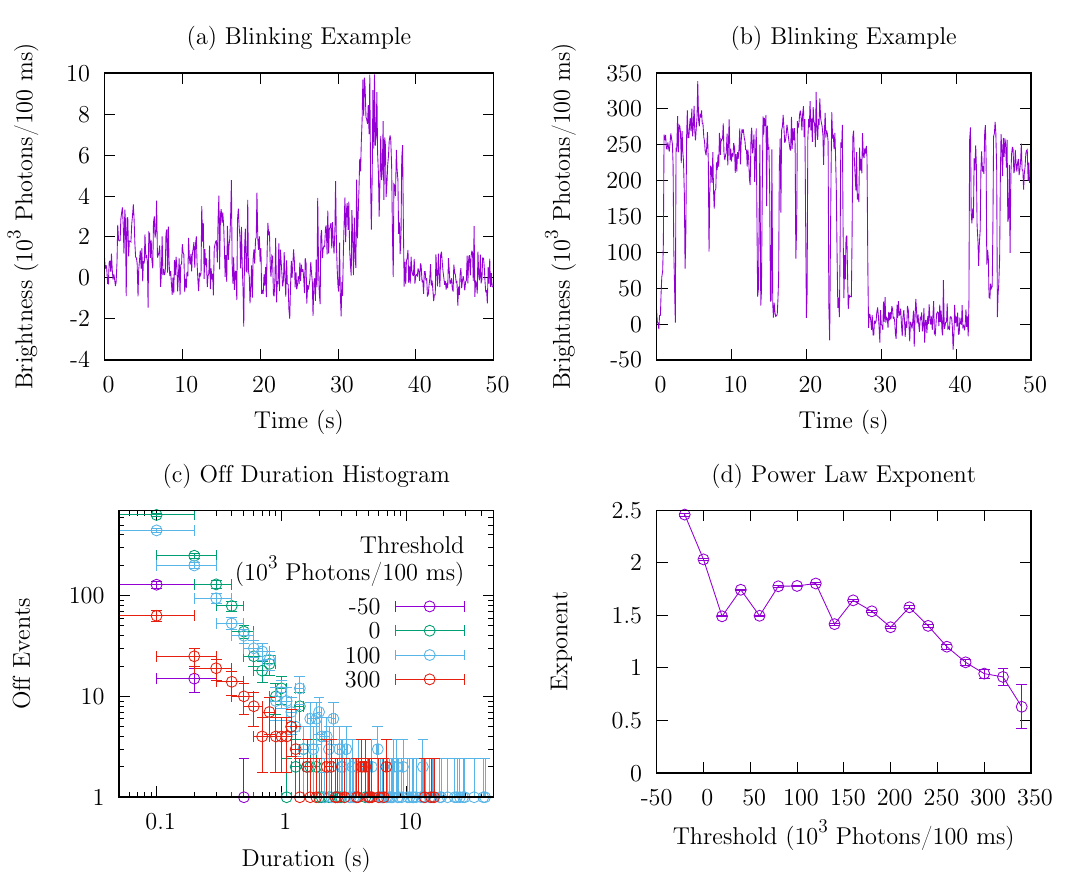}
    \caption{Examples of the interaction between student decisions and blinking results.  (a, b) Contrasting student-recorded single quantum dot photon trajectories.  (a) Quantum dots without shells in poly(methyl methacrylate) and hexadecylamine. (b) Core-shell-shell quantum dots in poly(methyl methacrylate) with only the surfactants/ligands present from the synthesis. (c) Authors' blinking off duration histograms generated from multiple student measurements under condition (b), for various brightness thresholds. (d) Authors' calculated power law exponent as a function of threshold for the same data set as (c).  The error bars indicate random error estimates.  The choice of threshold changes both systematic and random error.}
    \label{fig:threshold}
\end{figure}

Fig. \ref{fig:blink} shows an example video frame of single quantum dot luminescence. Students observe less than one layer of luminescent quantum dots.  The single quantum dot emission can be observed using the eyepieces of the microscope, with both the emission colour and blinking obvious.  Fig. \ref{fig:threshold}(a) and (b) are examples of the large variation in blinking behavior caused by students' choice of sample preparation conditions.  In (a), the quantum dots without shells are mostly off.  In (b), the core-shell-shell quantum dots are mostly on, with a larger peak brightness and much clearer on/off contrast. The photon trajectories obtained under each condition is consistent with literature measurements.\cite{gomez2006blinking}  The noise level is about \SI{3e3}{photons\per 100\, \milli\second}.

\sisetup{
fixed-exponent      = 4        ,
list-units          = brackets ,
range-units         = brackets ,
scientific-notation = fixed,
range-phrase = --
}
Example Fig. \ref{fig:threshold}(c) shows how the histogram changes as a function of the choice of threshold.  Fig. \ref{fig:threshold}(d) indicates that the power law exponent also depends on the selected threshold.  The region where the exponent is insensitive to the threshold, \SIrange{2e4}{20e4}{photons\per 100\, \milli\second}, spans an order of magnitude.  Students can, and occasionally do, use the interaction between the threshold and power law exponent to judge the quality of the data.  Threshold insensitivity suggests a replicable measurement of the exponent.  A high degree of sensitivity suggests unsuccessful classification of on and off states.  For example, in Fig. \ref{fig:threshold}(c--d), a negative threshold incorrectly classifies background noise as short-lived off states. The exponents reported by students ranged from 1.1 to 2.1 across 62 experiments, with a mean of 1.5 and standard deviation of 0.2.  Since the random error estimates are typically 0.01, students' choices explain most of the variation in the results.  These random error estimates include variation across both blinks and particles.
\section{Student Decision Making Results}
Our investigation of students' decision making was approved by the Monash University Human Research Ethics Committee.
\subsection{Student Laboratory Reports}
\begin{figure}
    \centering
    \includegraphics[width=.5\textwidth]{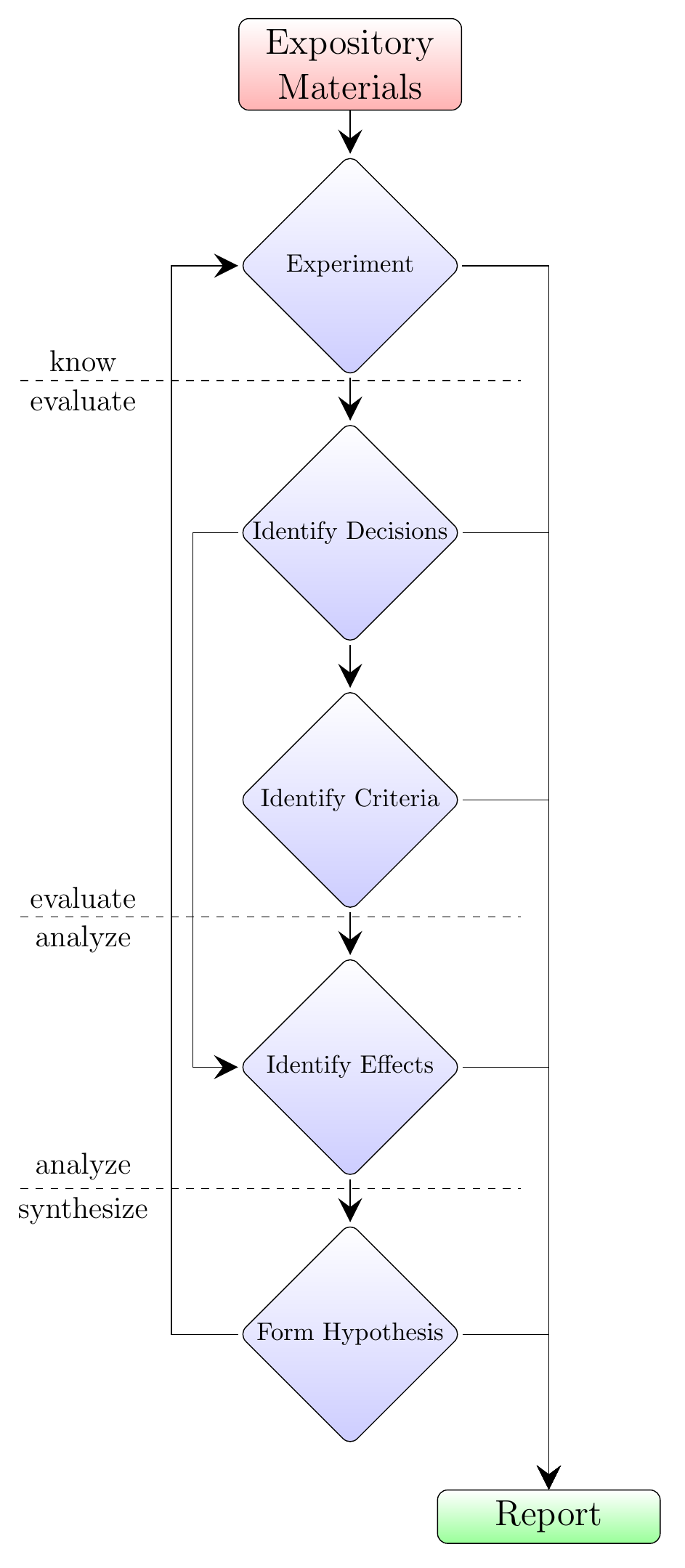}
    \caption{An illustration of students' thought processes, as inferred from their reporting of the decisions they made.}
    \label{fig:flow}
\end{figure}

We performed a retrospective, qualitative analysis of 45 students' reports to investigate what students thought about the decisions they made.  These students received no instructions to describe their decisions.  Fig. \ref{fig:flow} illustrates the cognitive processes we inferred from students' descriptions.  These congnitive processes contrast with the conventional scientific method and the experiment workflow.  Some students simply reported actions, such as ``A threshold was set.''  In these cases, it is unclear if students are unaware that they made a decision, if they chose to hide their decision-making because they believe decision making is unscientific, or if they lack the writing ability to clearly articulate their decisions.

Many students reported their decisions using decision words such as ``subjective,'' ``choose,'' ``judge,'' ``exclude,'' ``bias,'' or ``select.'' These words indicate the onset of critical thinking.  Students often wrote in the passive voice, leaving it open to our interpretation if the students believed they made the choice or that some external decider provided it.  We choose to interpret passive voice as indicating that the student made the choice.  Passive voice commonly indicates author actions in formal scientific writing.  Students may or may not have identified criteria for their choices, such as ``only the brighter dots or dots with more interesting blinking were chosen'' or ``two separate NPs overlaid in the same selection and were identifiable [for omission] by three distinct intensity levels.''  

Some students proceeded to state that their decisions had effects.  For example, ``\dots the subjectivity of the applied [brightness] threshold may have introduced errors into the data,'' ``results could have varied by the wavelength of laser used,'' or  ``errors \dots may be due to the subjective nature of choosing which spots appear to be due to a single or multiple emitters, potential biases when selecting which spots to monitor, subjectivity in the choice of the intensity above which the spots are considered `on' and below which they are considered `off' and the subjectivity in choosing a data fit range.''  These types of comments indicate that students understand their choices were important because they changed the results.  

Some students identified the dependent variable, such as, ``intensity.''  Only rarely did students identify an effect without the decision that caused it, such as, ``If there were overlapping quantum dots, then the intensity value may have increased.''  In this case, the student has not described their choice of dilution or their choice of image pixels as playing a role in the ``overlapping'' of quantum dots that were not spatially resolved in the microscope.  Identification of effects implies students have transitioned from the ``evaluate'' portion of the Bloom Taxonomy of Learning\cite{bloom} to the ``analyze'' portion.

As an additional level of complexity, some students formed specific hypotheses about the relationship between a decision and a dependent variable.  For example, ``When the threshold was too high, it meant that the nanoparticle[s] were in the `on' state too much which would have resulted in a less steep gradient.''  This contrasts with students who did not have any hypothesis about the direction of change in the dependent variable.  These students have entered the ``synthesize'' portion of critical thinking, which goes beyond our learning objectives.

Finally, a few students conducted an experimental test of their hypotheses, leading to a better experiment or better error estimates.  Examples include, ``If the sample was too dilute it became more difficult to locate the quantum dots, and therefore the experiment had to be repeated with a more concentrated solution,'' or ``moving the [brightness] threshold line by a small amount could change the final $\mu$ value by up to $\pm$ 0.05.''  $\mu$ refers to the power law distribution exponent, which is the main parameter describing blinking.
\subsection{Student Reflections}
\subsubsection{First Survey}
To investigate students' views of the decisions they made, we used a survey.  Fifteen students voluntarily participated by completing written free responses to four prompts.  The students' reflective comments were collected immediately after they finished analyzing their data with the provided software.

The first prompt probed students' knowledge of the decisions they made when excluding data.  In response to the question ``During the prac, how did you determine which data to use, at each stage of the analysis?'' 73\% of the participants identified at least one decision about excluding data.  The remaining students referred to authorities, including written instructions or instructors, or indicated that they guessed.  One student wrote, ``I used some data that in hindsight I probably shouldn't have,'' indicating that they learned more about data selection as they completed the data analysis.

The second prompt investigated students' decisions when setting the threshold, which is used to convert brightnesses to blink durations.  We asked, ``What  did you consider when setting the threshold for on and off?  How did you feel about the threshold selected?''  53\% of students were able to identify at least one criterion they used to decide the threshold.  For example, one student wrote ``I chose a threshold between on and off brightness...''  The criteria given varied.  The difficulty of identifying these criteria depends on the sample the student was investigating.  When measuring quantum dot cores, the on-state intensity is not sufficient to allow students to clearly identify a particular brightness level as indicating an on-state.

We asked, ``What sources of experimental bias could be in the experiment?'' to find out if students could identify their choices as sources of bias.  93\% of students identified at least one source of bias.  80\% of students identified at least one source of bias that was related to their decisions.  Overall, every student participant indicated that they were thinking about their decisions in at least one of three ways: identification of a decision to exclude data, identification of criteria for setting a threshold, or identification of bias related to a decision.  

The final question probed students' analytical skill.  It was ``How could bias change your results?''  Students had a lot of difficulty with this question.  While 60\% of students identified a potential effect of bias, only 20\% suggested that bias might change the value of the power law exponent, $\mu$.  None of those students successfully identified a relationship between a bias and the distribution of off-state durations. Two factors can contribute to this failure:  First, students had not compared $\mu$ measurements across different sample conditions at this stage in the activity.  Therefore they may not have been prompted to realize $\mu$ was a result.  Second, students may have inadequate mathematical preparation to understand power laws.

\subsubsection{Second Survey}
At the end of students' lab reports, a second survey was conducted.  The same fifteen students participated.  Students received a grade for answering the survey, but the graders were blind to the student's consent to participate in the research.  First, we asked ``How did you decide which data to use in your report?'' At this stage, 93\% of students identified at least one decision making criterion.  None of the students referred to authorities.  This indicates a pronounced increase in students' awareness of their own decisions between the first and second survey.  This may indicate students were learning as they prepared their reports, or it may indicate greater student engagement with the second survey.

To probe students' views of the relative importance of sample preparation decisions and data analysis decisions, we asked ``How might your measurements be different from those of other students?''  80\% of students identified analysis choices as the reason for differences between measurements.  20\% of students identified sample preparation choices.  Most of that 20\% gave both reasons.

To follow up on that question, we asked ``Were the differences between samples caused by bias or were they caused by deliberate differences in the way samples were prepared? How could you tell?''  53\% of students identified bias as the cause of differences.  Analysis choices, inadvertent variation, and random errors were included among the examples of bias listed by students.  47\% of students used internal comparisons to identify bias or lack of bias.  For example, ``the replicates of the same conditions that were analysed by different people varies [sic] drastically'' or ``samples that were replicated by different people, as the results for those were reasonably close, and noticeably different when compared to other samples preparations.''  

Finally, we probed students' critical thinking about the literature with ``How did you decide if the results of the prac were the same or different from literature reports?''  50\% of students identified methodological choices as the reason they believed their results were different from the literature.   33\% identified correlations between their data and the literature.  20\% of students used internal comparisons to argue that their results must be different from the literature because students' results were inconsistent.  For example:  ``Some of the data obtained was quite conflicting which indicates that at least parts of the data are incorrect.''   None of the students mentioned using error estimates.  These survey responses give further evidence of students' critical thinking and awareness of the importance of methodological decisions.

\subsection{Focus Groups}
After students completed their reports, we conducted two focus groups\cite{vaughn1996focus} with three and four student participants.  As a warm-up, we asked students about the nature of laboratory instruction.  Interestingly, students did not form a consensus about what laboratory instruction is, except that it involves a task.  Next, students were prompted to discuss the specific decisions they made at various stages of the experiment they performed.  As in the reports and written reflections, students identified multiple decisions they made and discussed how those decisions related to the results.  We asked students about their views on decision making and the scientific process.  The participants indicated that they believed decision making is integral to the process of science: 
\vspace{1em}

\begin{minipage}{\textwidth}
\begin{dialogue}
  \direct{Group 1}
  \speak{Moderator} So how does decision making fit into the scientific process?
  \speak{Student F} It's fundamental to the science process.  You have to decide what you wanna look at and how you're going to look at and what implications your method has on what you're trying to achieve as well.
  \speak{Student H} Very integral. My project's actually on [redacted] so yeah because they, in industry or independent research there's like there isn't always gonna be that kind of, you know, guide the we need the students or like we as students need to have these decision making things down pat so you can ask your own questions and you can kind of you know have some form of independence.
  \speak{Student G} Decision making to science as a whole is very integral. So what methods you're going to use, what you're going to do, and stuff like that. But in terms of the undergraduate labs here at [redacted], it's very much do this, do that, use this, use that, then you try and let you choose some things like perhaps choosing [unclear] choosing what materials we should be using but its a lot very structured you just get that, and then you just analyze this.
\end{dialogue}
\end{minipage}
\vspace{1em}

\begin{minipage}{\textwidth}
\begin{dialogue}
  \direct{Group 2}
  \speak{Moderator} So, is decision making part of the scientific process?
  \speak{Student B} I think so its under the yeah.
  \speak{Student D} I'd say so.
  
  \direct{Laughter}
  \speak{Student B} Everything's decision making I feel like.
  \speak{Student T} I'd say it is, but we skipped a lot the decision making ourselves because it was over and done. So we came kind of later into the into the scientific process. That's all I want to say.
  \speak{Moderator} So do the decisions scientists make change our understanding of science?
  \speak{Student D} Yes, I guess they can.
  
  \direct{Laughter}
  \speak{Student T} [unclear] decide to study what they study they probably wouldn't be studying they would go about it.
  \speak{Student D} Someone could decide not to use less favorable data.  It's not good to do that but if you're deciding not to do that then you're missing out on results that may be different from what you find in the end.
  \speak{Moderator} O.K., I see some nodding.
\end{dialogue}
\end{minipage}
\vspace{1em}

These discussions indicate that the participants understand the importance of decision making to science.  They gave contrasting reasons, including personal autonomy, lack of background information, scientific integrity, and need for a topic, for why decision making is part of the scientific process.  None of the participants took the positivist position that scientific knowledge is based purely on evidence, exclusive of the choices of the investigator.  We find that the participants value decision making, which is a type of critical thinking, in scientific contexts.

\section{Conclusions}

In conclusion, analyzing the stochastic luminescence intermittencies of quantum dots as part of laboratory instruction develops students' critical thinking skills.  Students are required to make several choices in order to record blinking videos and reduce the videos to a single parameter.  Not all these choices have unambiguously correct answers.  Therefore, the experiment presents beneficial challenges to high-performing students, while still easily and consistently producing measurements.  The content is interdisciplinary, relating chemistry, quantum mechanics, and statistics.

Students identified that they made decisions, used criteria, and had bias in their measurements.  None of these observations are consistent with positivist epistemology.  They are also incompatible with the belief in epistemic authority, which asserts that experts are the source of truth. \cite{raviv2003teachers,hornikx2011epistemic,njegovan2011characteristics,blue2018assessing}  We find that, while students hold a range of beliefs about the origin of scientific knowledge, investigating quantum dots provides students with experience supporting constructivist epistemology.

\section{Associated Content}
\begin{sloppypar}The Supporting Information is available on the ACS Publications website at DOI: 10.1021/acs.jchemed.XXXXXXX.
\end{sloppypar}

Student handout with activity protocol, survey protocol (PDF).

\section{Acknowledgements}
We gratefully acknowledge the technical assistance of Rosalind Cox, Esther Miriklis, Riley Hargreaves, Tich-Lam Nguyen, Paul Mulvaney, and Anum Nisar.  We acknowledge Tina Overton for helpful discussions.  We also acknowledge support from Michael Grace.  
A. M. F. and L. F. acknowledge support by the Australian Research Council (ARC) via the ARC Centre of Excellence in Exciton Science (CE170100026).
T. B. acknowledges support from the Australian Research Council (DP170104477).
\bibliography{prac}
\end{document}